\documentstyle[twocolumn,epsfig,prl,aps]{revtex}

\begin{document}
\draft
\title{Experimental observation of nonlinear Thomson scattering}
\author{Szu-yuan Chen, Anatoly Maksimchuk \& Donald Umstadter}
\address{Center for Ultrafast Optical Science, University of Michigan, Ann Arbor, MI\\
48109, USA}
\date{\today}
\maketitle
\pacs{}

\preprint{version 2.1}
\narrowtext

{\bf A century ago, J. J. Thomson\nocite{Thomson}$^{1}$ showed that the
scattering of low-intensity light by electrons was a linear process (i.e.,
the scattered light frequency was identical to that of the incident light)
and that light's magnetic field played no role. Today, with the recent
invention of ultra-high-peak-power lasers\nocite{Mourou}$^{2}$ it is now
possible to create a sufficient photon density to study Thomson scattering
in the relativistic regime. With increasing light intensity, electrons
quiver during the scattering process with increasing velocity, approaching
the speed of light when the laser intensity approaches 10$^{18}$ W/cm$^{2}$.
In this limit, the effect of light's magnetic field on electron motion
should become comparable to that of its electric field, and the electron
mass should increase because of the relativistic correction. Consequently,
electrons in such high fields are predicted to quiver nonlinearly, moving in
figure-eight patterns, rather than in straight lines, and thus to radiate
photons at harmonics of the frequency of the incident laser light\nocite
{Vachaspati, Brown,Sarachik,Castillo,Esarey,Hartemann,Hartemann2}$^{3-9}$,
with each harmonic having its own unique angular distribution\nocite
{Sarachik,Castillo,Esarey}$^{5-7}$. In this letter, we report the first ever
direct experimental confirmation of these predictions, a topic that has
previously been referred to as nonlinear Thomson scattering\nocite{Esarey}$%
^{7}$. Extension of these results to coherent relativistic harmonic
generation\nocite{Esarey2,Esarey3}$^{10,11}$ may eventually lead to novel
table-top x-ray sources.}

In this experiment, we used a laser system that produces 400-fs-duration
laser pulses at 1.053-$\mu $m wavelength with a maximum peak power of 4 TW.
The 50-mm diameter laser beam was focused with an f/3.3 parabolic mirror
onto the front edge of a supersonic helium gas jet. The focal spot is
consisted of a 7-$\mu $m FWHM Gaussian spot (containing 60 $\%$ of the total
energy) and a large ($>$ 100 $\mu $m) dim spot. The helium gas was fully
ionized by the foot of the laser pulse. A half-wave plate was used to rotate
the axis of linear polarization of the laser beam in order to vary the
azimuthal angle ($\phi $) of observation. We define $\theta =0^{\circ }$ as
along the direction opposite to that of the laser propagation and $\phi
=0^{\circ }$ as along the axis of linear polarization. In a linearly
polarized laser field, electrons move in a figure-eight trajectory lying in
the plane defined by the axis of linear polarization and the direction of
beam propagation.

While the observation of harmonics in laser-plasma (or electron beam)
interactions has been made by several groups\nocite
{Meyer,Basov,Malka,Englert,Bula}$^{12-16}$, that alone is insufficient to
unambiguously identify nonlinear Thomson scattering and its underlying
dynamics. Several other mechanisms might generate continuum or harmonics
under our experimental conditions, and, therefore, need to be isolated and
discriminated from the signal generated by nonlinear Thomson scattering: (1)
continuum generated from self-phase modulation of laser beam in gas, (2)
harmonics generated from atomic nonlinear susceptibility of gas or,
especially, from the ionization process\nocite{Brunel}$^{17}$, (3) continuum
generated from (a) (relativistic) self-phase modulation of laser pulse in
the plasma, or from (b) electron-electron bremsstrahlung and electron-ion
bremsstrahlung, and (4) harmonics generated from the interaction of laser
pulses with a transverse electron-density gradient\nocite{Malka}$^{14}$.

The main focal spot of the laser pulse undergoes relativistic-ponderomotive
self-channeling when high laser power and gas density are used\nocite{Chen}$%
^{18}$. Side imaging ($\theta $ = 90$^{\circ }$) of the 1st harmonic light
(at the laser frequency) from nonlinear Thomson scattering shows that the
laser channel has a diameter of $<$10 $\mu $m FWHM. However, interferograms 
\nocite{Chen}$^{18}$ show that the diameter of the plasma column is about
100-200 $\mu $m, which is created by the wings with intensities $>10^{15}$
W/cm$^{2}$ (the ionization threshold). Therefore, the light generated from
laser-gas interaction should be observed to originate from the entire region
of plasma, rather than from the narrow laser channel. Results of side
imaging ($\theta $ = 90$^{\circ }$ and arbitrary $\phi $) of the 2nd and 3rd
harmonics using a matching interference filter (10 nm bandwidth) show that
the signal is emitted only from the narrow laser channel. In addition, the
images of the harmonics have spatial distributions similar to the images of
the 1st harmonic light, and their profiles vary in the same way as the laser
power and gas density are changed. This rules out the possibility that the
harmonic signal observed in the side images is a result of laser-gas
interaction ((1) and (2)).

\begin{figure}
\centering{\epsfig{file=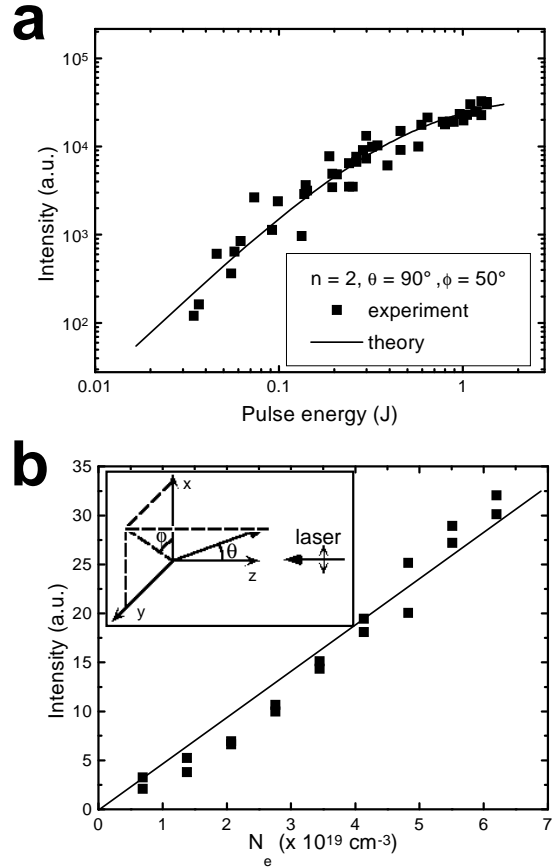,width=1.0\columnwidth}}
\caption{Intensity of the 2nd harmonic light at $\protect\theta = 90^{\circ}$%
, $\protect\phi = 50^{\circ}$ ({\bf a}) as a function of laser pulse energy
at $6.2\times 10^{19}$-cm$^{-3}$ electron density and ({\bf b}) as a
function of plasma electron density at 0.8-J laser pulse energy. (For 1-J
laser pulse energy, the laser intensity is $4.4\times 10^{18}$ W/cm$^2$ and $%
a_0$ is 1.88.) Each data point represents the result of a single laser shot.
The theoretical predictions for zero drift velocity are plotted in solid
lines for comparison. The only fitting parameter is just a constant for
normalization in all figures. The laser pulse energy refers to the energy in
the main focal spot. The inset shows the coordinate system used.}
\label{powden}
\end{figure}

According to theory\nocite{Sarachik,Castillo,Esarey}$^{5-7}$, the harmonic
signal generated from nonlinear Thomson scattering should have two important
features: (1) it is linearly proportional to the electron density because it
is an incoherent single electron process (the harmonics generated from a
collection of electrons interfere with each other destructively, leaving
only an incoherent signal, which is equal to the single-electron result
multiplied by the total number of electrons which radiate), and (2) it
increases roughly as $I^{n}$, where $n$ is the harmonic number, and
gradually saturates when $a_{0}$ is on the order of unity\nocite{Esarey}$%
^{7} $, where $a_{0}=eE/m_{0}\omega _{0}c=8.5\times 10^{-10}\lambda $[$\mu $%
m]$I^{1/2}$[W/cm$^{2}$] is the normalized vector potential, $E$ is the
amplitude of laser electric field, and $I=cE^{2}/8\pi $ is the laser
intensity. These are characteristically different from the behavior of any
other mechanisms. For instance, bremsstrahlung radiation should be
proportional to the square of gas density ($N_{e}\cdot N_{e}$ or $N_{e}\cdot
N_{i}$). In this experiment, the intensity of the harmonic signal was
determined from the peak intensity or the average intensity of the images of
harmonics, when it was plotted as a function of the observing angle, gas
density and laser power. Both showed the same variations. Figure\thinspace 
\ref{powden} shows the variation of the 2nd harmonic signal as a function of
laser power and plasma (electron) density. The experimental results show a
reasonable fit with the theoretical predictions. The 1st and 3rd harmonics
show the same match with the theory.

\begin{figure}
\centering{\epsfig{file=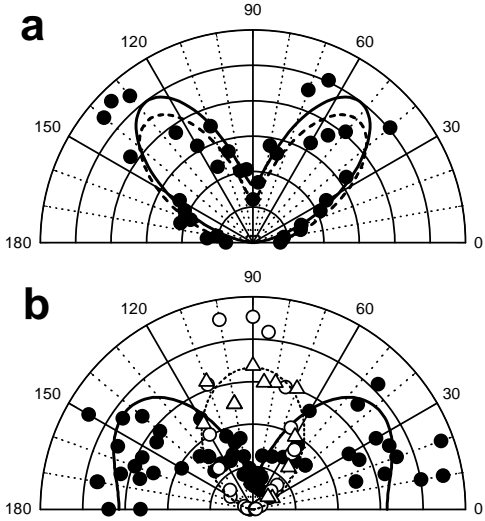,width=1.0\columnwidth}}
\caption{Polar plots of the intensity of the 2nd harmonic light as a function of
azimuthal angle ($\protect\phi$, in degrees) for 0.8-J pulse energy and $%
6.2\times 10^{19}$-cm$^{-3}$ electron density at ({\bf a}) $\protect\theta =
90^{\circ}$ and ({\bf b}) $\protect\theta = 51^{\circ}$. The intensity is in
arbitrary units. The solid circles represent the experimental data. The
solid and dash lines represent the theoretical results for zero and nonzero
drift velocity, $v = 0.2 c$, in the laser propagation direction,
respectively. The open circles and triangles represent the experimental data
for the 1st harmonic signal taken at two different runs under the same
conditions. The dotted line represents the theoretiocal result. Its dipole
radiation pattern (peaked at $\protect\phi =$90$^{\circ }$) confirms that
there is no depolarization effect in the plasma and that the collective
effect of plasmas on the angular pattern is not significant (at least
outside of a narrow cone along the axis of laser propagation). Although the
data for $\protect\phi = 180^{\circ} \sim 360^{\circ}$ are not plotted, it
should be just a mirror image of the data for $\protect\phi = 0^{\circ} \sim
180^{\circ}$ because of the intrinsic symmetry of the laser field, as
predicted theoretically. Such symmetry has been verified in the experiment.}
\label{plon2f90}
\end{figure}

\begin{figure}
\centering{\epsfig{file=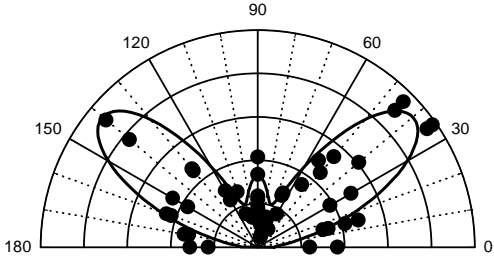,width=1.0\columnwidth}}
\caption{Polar plot of the intensity of the 3rd harmonic light as a function of
azimuthal angle ($\protect\phi$, in degrees) at $\protect\theta = 90^{\circ}$
for 0.8-J pulse energy and $6.2\times 10^{19}$-cm$^{-3}$ electron density.
The intensity is in arbitrary units. The solid circles represent the
experimental data. The solid line represents the theoretical result for zero
drift velocity. The angular patterns of harmonics should not be sensitive to
variation of laser intensity, as expected from theory and checked in the
experiment. This is crucial to the success of our measurements because it
alleviates the error caused by fluctuation of laser intensity.}
\label{plon3f90}
\end{figure}

Although the above two observations are consistent with nonlinear Thomson
scattering as the source of the harmonic signal, the observation of the
unique angular patterns is necessary in order to prove that the detailed
dynamics of nonlinear Thomson scattering are indeed the same as the
theoretical prediction. Figure\thinspace \ref{plon2f90}(a) shows the $\phi $
-dependence of the 2nd harmonic signal at $\theta =$ 90$^{\circ }$. The
experimental results match qualitatively with the theoretical prediction,
both having a quadrupole-type radiation pattern, which is characteristically
different from the dipole pattern for other mechanisms (1)-(4), and linear
Thomson scattering. Other measurements such as the $\phi $-dependence of the
2nd harmonic light at $\theta =$ 51$^{\circ }$ (an ``anti-dipole'' pattern),
shown in Fig.\thinspace \ref{plon2f90}(b), and the $\phi $-dependence of the
3rd harmonic light at $\theta =$ 90$^{\circ }$ (a ``butterfly'' pattern),
shown in Fig.\thinspace \ref{plon3f90}, were also made, all showing
reasonable matches between the experimental data and the theoretical
predictions. Such angular radiation patterns directly prove that electrons
do indeed oscillate with figure-eight trajectories in an intense
(relativistic) laser field. The angular pattern of the 1st harmonic light
(linear component) of nonlinear Thomson scattering is also included in
Fig.\thinspace \ref{plon2f90}(b) for comparison.

\begin{figure}
\centering{\epsfig{file=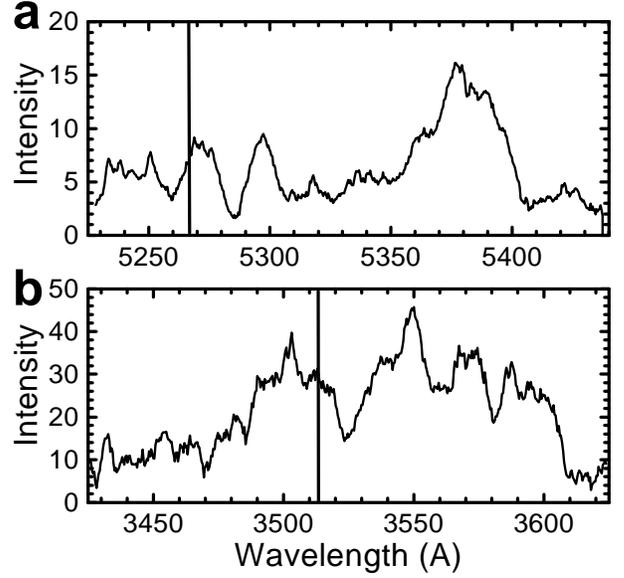,width=1.0\columnwidth}}
\caption{Spectra of the ({\bf a}) 2nd and ({\bf b}) 3rd harmonics at $%
\protect\theta = 90^{\circ}$, $\protect\phi = 50^{\circ}$ for 0.8-J pulse
energy and $6.2\times 10^{19}$-cm$^{-3}$ electron density. Vertical lines
indicate the wavelengths of the unshifted 2nd and 3rd harmonics. The
intensities are ploted in arbitrary units. The spectra do not change with
variation of $\protect\phi$ at any specific $\protect\theta$, so the angular
distributions measured are not affected by the choice of filter bandwidth.}
\label{spectra}
\end{figure}

Measurements of the spectra of the harmonics show that each of the spectra
of 2nd and 3rd harmonics contains a peak at roughly the harmonic wavelength
and a red-shifted broader peak, as shown in Fig.\thinspace \ref{spectra}.
The red-shifted broader peaks are believed to be part of the harmonic
spectra generated by nonlinear Thomson scattering, because they vary in
amplitude proportionally with the corresponding unshifted harmonic signals
when the gas density and the laser power are changed. It was expected that the
spectra of harmonics should be broadened tremendously for electrons in a
high-fluid-velocity plasma wave\nocite{Castillo}$^{6}$. A
fast-phase-velocity electron plasma wave (with a maximum fluid velocity of
as large as $\sim $0.2 $c$, where $c$ is the speed of light in vacuum)
excited by stimulated Raman forward scattering\nocite{LeBlanc}$^{19}$ was
observed in this experiment at high laser power and gas density. However,
the fact that the spectral distribution of the harmonics was not observed to
change significantly with variation of gas density and laser power, when the
plasma wave amplitude was, indicates that such spectral structure has
nothing to do with the collective drift motion of electrons in the plasma
waves. Although the angular radiation patterns of the harmonics could also
be affected by such a 0.2 $c$ fluid-velocity oscillation, the changes are
not significant enough (compare the solid and dash lines in Fig.\thinspace 
\ref{plon2f90}(a)) to be identified from the experimental data\nocite{Esarey}%
$^{7}$. In other words, all measurements done in this experiment match
qualitatively with the prediction of incoherent nonlinear Thomson scattering
of electrons without drift motion; the results appear not to be affected by
the existence of plasma waves, probably due to destructive coherent
interference. The absolute scattering efficiency is measured to be $8\times
10^{-4}$ and $1\times 10^{-4}$ photons per electron per pulse for the 2nd
and 3rd harmonics (including both the unshifted and red-shifted spectral
components), respectively, at $\theta =90^{\circ }$, $\phi =50^{\circ }$,
for an angle of collection of $7\times 10^{-3}$ steradians. These numbers
match reasonably well with the theoretical predictions for incoherent
nonlinear Thomson scattering, which are $8\times 10^{-4}$ and $5\times
10^{-4}$, respectively.

In summary, the results reported here confirm for the first time several
predictions of relativistic electrodynamic theory, which were formulated
forty years ago, coincident with the invention of the laser. As predicted 
\nocite{Esarey}$^{7}$, a century-old fundamental ``constant,'' the Thomson
cross-section, is now shown to depend on the strength of light.

\begin{center}
{\large REFERENCES}
\end{center}

\begin{enumerate}
\item  \nocite{Thomson} Thomson, J. J. Conduction of electricity through
gases. Cambridge University Press, Cambridge (1906).

\item  \nocite{Mourou} Maine, P. {\em et al.} Generation of ultrahigh peak
power pulses by chirped pulse amplification. IEEE J. Quantum Electron. {\bf %
24}, 398-403 (1988).

\item  \nocite{Vachaspati} Vachaspati Harmonics in the scattering of light
by free electrons. Phys. Rev. {\bf ~128}, 664-666 (1962).

\item  \nocite{Brown} Brown, L. S. \& Kibble, T. W. B. Interaction of
intense laser beams with electrons. Phys. Rev. {\bf ~133}, A705-A719 (1964).

\item  \nocite{Sarachik} Sarachik, E. S. \& Schappert, G. T. Classical
theory of the scattering of intense laser radiation by free electrons. Phys.
Rev. D {\bf ~1}, 2738-2753 (1970).

\item  \nocite{Castillo} Castillo-Herrera, C. I. \& Johnston, T. W.
Incoherent harmonic emission from strong electromagnetic waves in plasmas.
IEEE Trans. Plasma Sci. {\bf 21}, 125-135 (1993).

\item  \nocite{Esarey} Esarey, E., Ride, S. K. \& Sprangle, P. Nonlinear
Thomson scattering of intense laser pulses from beams and plasmas. Phys.
Rev. E {\bf ~48}, 3003-3021 (1993).

\item  \nocite{Hartemann} Hartemann, F. V. \& Luhmann, N. C. Jr. Classical
electrodynamical derivation of the radiation damping force. \prl {\bf~74},
1107-1110 (1995).

\item  \nocite{Hartmann2} Hartemann, F. V. High-intensity scattering
processes of relativistic electrons in vacuum. Phys. Plasmas {\bf ~5},
2037-2047 (1998).

\item  \nocite{Esarey2} Esarey, E. {\em et al.} Nonlinear analysis of
relativistic harmonic generation by intense lasers in plasmas. IEEE Trans.
Plasma Sci. {\bf ~21}, 95-104 (1993).

\item  \nocite{Esarey3} Esarey, E. \& Sprangle, P. Generation of stimulated
backscattered harmonic generation from intense-laser interactions with beams
and plasmas. Phys. Rev. A {\bf ~45}, 5872-5882 (1992).

\item  \nocite{Meyer} Meyer, J. \& Zhu, Y. Second harmonic emission from an
underdense laser-produced plasma and filamentation. Phys. Fluids {\bf ~30},
890-895 (1987).

\item  \nocite{Basov} Basov, N. G. {\em et al.} Investigation of 2$\omega _0$%
-harmonic generation in a laser plasma. Sov. Phys. JETP {\bf ~49}, 1059-1067
(1979).

\item  \nocite{Malka} Malka, V. {\em et al.} Second harmonic generation and
its interaction with relativistic plasma waves driven by forward Raman
instability in underdense plasmas. Phys. Plasmas {\bf ~4}, 1127-1131 (1997).

\item  \nocite{Englert} Englert, T. J. \& Rinehart, E. A. Second-harmonic
photons from the interaction of free electrons with intense laser radiation. %
\pra {\bf~28}, 1539-1545 (1983).

\item  \nocite{Bula} Bula, C. {\em et al.} Observation of nonlinear effects
in Compton scattering. \prl {\bf~76}, 3116-3119 (1996).

\item  \nocite{Brunel} Brunel, F. Harmonic generation due to plasma effects
in a gas undergoing multiphoton ionization in the high-intensity limit. %
\josab {\bf 7}, 521-526 (1990).

\item  \nocite{Chen} Chen, S.-Y. {\em et al.} Evolution of a plasma
waveguide created during relativistic-ponderomotive self-channeling of an
intense laser pulse. \prl {\bf 80}, 2610-2613 (1998).

\item  \nocite{LeBlanc} Le Blanc, S. P. {\em et al.} Temporal
characterization of a self-modulated laser wakefield. \prl {\bf 77},
5381-5384 (1996).
\end{enumerate}

{\bf Acknowledgements} This work was supported by U. S. National Science
Foundation and the Division of Chemical Sciences, Office of Basic Energy
Sciences, Office of Energy Research, U.S. Department of Energy. The authors
would also like to thank G. Mourou, R. Wagner and X.-F. Wang for their
useful discussions.

\end{document}